\documentclass[journal]{spie}

\usepackage{cite}
\usepackage{graphicx}
\usepackage{url}

\usepackage[switch]{lineno}

\usepackage{amssymb}
\usepackage{latexsym}
\usepackage{appendix}

\usepackage{color}

\newcommand{\edit}[1]{\textcolor{black}{#1}}
\newcommand{\responseedit}[1]{\textcolor{black}{#1}}
\newcommand{\minoredit}[1]{\textcolor{black}{#1}}

\newcommand{\omitting}[1]{}

\usepackage{newfloat}
\DeclareFloatingEnvironment[name={Figure A}]{suppfigure}

\graphicspath{{Figures/}}

\begin{document}

\title{Segmentation of \edit{MRI head anatomy using} deep volumetric networks and multiple spatial priors}

\author{Lukas~Hirsch,
        Yu~Huang\thanks{Both authors contributed equally.},
        and~Lucas~C~Parra \thanks{This work was supported in part by NIH with Grants R01MH111896, R01NS095123,  R21NS115018 and R44NS092144.} 
\thanks{L. Hirsch and L. Parra are with the Department
of Biomedical Engineering, City College New York, New York City,
NY, 10031 USA e-mail: parra@ccny.cuny.edu. Y. Huang is with the Department
of Radiology, Memorial Sloan Kettering Cancer Center, New York City,
NY, 10065 USA.}
}

\maketitle

\begin{abstract}

Purpose: Conventional automated segmentation of the head anatomy in MRI distinguishes different brain and non-brain tissues based on image intensities and prior tissue probability maps (TPM). This works well for normal head anatomies, but fails in the presence of unexpected lesions. Deep convolutional neural networks leverage instead spatial patterns and can learn to segment lesions, but often ignore prior probabilities.

Approach: We add three sources of prior information to a three-dimensional convolutional network, namely, spatial priors with a TPM, morphological priors with conditional random fields, and spatial context with a wider field-of-view at lower resolution. We train and test these networks on 3D images of 43 stroke patients and 4 healthy individuals which have been manually segmented. 

Results: We demonstrate the benefits of each sources of prior information, and we show that the new architecture, which we call Multiprior network, \responseedit{improves} the performance of existing segmentation software, such as SPM, FSL, and DeepMedic \responseedit{for abnormal anatomies}. \minoredit{The relevance of the different priors was compared and the TPM was found to be most beneficial}. The benefit of adding a TPM is generic in that it can boost the performance of established segmentation networks such as the DeepMedic and a UNet. We also provide an out-of-sample validation and clinical application of the approach on an additional 47 patients with disorders of consciousness. We make the code and trained networks freely available.

Conclusions: Biomedical images follow imaging protocols that can be leveraged as prior information into deep convolutional neural networks to improve performance. The network segmentations match human manual corrections performed in 3D, and are \responseedit{comparable} in performance to human segmentations obtained from scratch in 2D \responseedit{for abnormal brain anatomies}.

\end{abstract}

\omitting{
}


\section{Introduction}

Clinical and basic research require segmentation of magnetic resonance images (MRI) of human heads, including abnormal anatomies such as tumors or lesions. In the case of brain tumors, it is helpful to measure the tumor volume across successive scans to monitor tumor growth or the response to treatment \cite{\omitting{konukoglu_monitoring_2008,}bauer_survey_2013}. In the case of a stroke, lesion studies can provide important insights on brain function \cite{alexander_correlating_2010\omitting{crinion_neuroimaging_2013}}. \edit{In patients with traumatic brain injury, volumetric analysis can provide diagnostic information such as hydrocephalus \cite{miskin_diagnosis_2017}.} 
When analyzing large populations this is only possible with automated segmentation  \cite{llado_segmentation_2012\omitting{pustina_automated_2016, mckinley_fully_2017, seghier_lesion_2008}}. Finally, transcranial electric stimulation techniques rely on accurate head segmentations of individual subjects \cite{dmochowski_targeted_2013}. Such a segmentation needs to capture not only the brain and lesion but also the cerebrospinal fluid, ventricles, air cavities, skull, etc \cite{datta_individualized_2011}. It is not feasible to do this manually even on a moderate number of cases.  

A number of tools have been developed to automate the task of segmenting the brain. This includes algorithms that are part of neuroimaging software packages such as SPM \cite{ashburner_unified_2005} and FSL \cite{smith_advances_2004}. These algorithms traditionally distinguish different tissues based on the brightness of voxels. For instance, in T1-weighted MRI white matter is bright, gray matter is gray, and surrounding CSF and skull are black. Segmentation also relies on prior information on the type of tissues that can be expected at different location of the head, e.g., with very high probability the surface of the head is skin. These probabilities are derived from manual segmentations of a large number of heads and are stored as tissue probability maps (TPM) \cite{ashburner_unified_2005}. TPMs are available only for normal anatomies. Therefore, traditional algorithms that are based on these TPMs work well for subjects with normal anatomy, but often fail in the presence of lesions. For example, in chronic stroke patients, areas that typically contain brain (bright) are filled with cerebrospinal fluid (dark in T1 images). This leads to ambiguities that confuse the traditional algorithms and can result in errors that extend beyond the lesions. As a result, MRIs from stroke patients require manual correction of the automated segmentations, despite efforts to improve on these traditional tools \cite{pustina_automated_2016, mckinley_fully_2017, seghier_lesion_2008}.

A breakthrough was achieved recently with deep convolutional networks that can identify tissues based on complex three-dimensional intensity patterns, instead of relying on single intensity values. For example, DeepMedic is a 3D convolutional network that achieves good segmentation of stroke lesions or brain tumors \cite{kamnitsas_efficient_2017}. \omitting{QuickNAT is a \edit{2D convolutional network that was trained to quickly parcellate the brain into different structures} \cite{guha_roy_quicknat:_2018}.} Convolutional networks can learn to identify complex features while limiting the number of parameters to be learned. This is accomplished by making feature extraction invariant to location. Yet, location is important in many medical imaging tasks. To take location into account one can provide the coordinates of each voxel as input to the network \cite{zikic_encoding_2014,ghafoorian_location_2017, wachinger_deepnat:_2018, guha_roy_quicknat:_2018,rachmadi_segmentation_2018, novosad_accurate_2019}. \omitting{The network can then learn which tissue is most likely to be found at particular locations. \edit{Another approach is to train different networks for different locations of the brain \cite{huo_3d_2019}.}} An alternative is to provide \edit{prior} probabilities explicitly as input using traditional TPMs \cite{zikic_encoding_2014,ghafoorian_location_2017, kushibar_automated_2018, yue2019cardiac}. This is the approach we adopt in this work.

Another important factor guiding medical segmentation is shape or morphology of a region. One approach has been to parameterize morphology using auto-encoders \cite{oktay_anatomically_2018, Dalca_2018_CVPR, larrazabal_anatomical_2019}. \omitting{When applied to existing segmentations, auto-encoders learn to represent the possible segmentations with fewer parameters. By forcing a neural network to use this lower-dimensional representations the network is biased to produce segmentations that are more likely, given these prior existing segmentations.} Another approach is to use conditional random fields (CRF), which implement morphological constraints \cite{chen_deeplab:_2018, kamnitsas_efficient_2017}. For instance, one might a-priori disallow segmentations where the brain touches the skin. Here we implement such morphological constraints as tissue-specific penalties in a fully connected 3D conditional random field. 

Spatial prior information can also be derived from context in an image. For instance, when segmenting breast tumors, it helps to know if a given voxel is part of the breast. This can be implemented by first identifying the breast in an image, and then providing this as input for the breast tumor segmentation \cite{zhang_hierarchical_2019}. We follow the approach of DeepMedic \cite{\omitting{kamnitsas_deepmedic_2016,} kamnitsas_efficient_2017} \edit{and VoxResNet \cite{chen_voxresnet:_2018}} where the network learns  contextual information from a wider field-of-view (FOV) at lower resolution.

In total, we implement three spatial priors: location priors with a TPM, neighborhood priors with a CRF, and context prior with a wider FOV, \edit{which are combined here for the first time in a deep network architecture.}  We therefore call this new deep, three-dimensional network, Multiprior. The resulting network is fully trainable, including the prior probabilities. Thus, the Multiprior architecture represents a convolutional network with learnable spatial memories. These memories have a simple interpretation, and can be manually instantiated based on prior spatial knowledge.  

In this work we \edit{focus on the task of segmenting the whole head in subjects with abnormal brain anatomy, which remains a challenging problem. \omitting{We are not concerned with normal anatomy, or a finer parcellation within the brain, for which there has been already significant progress using deep network techniques \cite{mehta_brainsegnet_2017,li_compactness_2017,chen_voxresnet:_2018,guha_roy_quicknat:_2018,huo_3d_2019}.}}  We train networks on available manual segmentations of stroke patients and normal subjects. Based on this \edit{reference} we objectively judge the benefits of adding prior information and spatial context. We then compare \edit{the results} to human expert segmentations\edit{, as well as conventional and state of the art segmentation methods.} The utility of the technique is demonstrated on a separate data set of patients with disorders of consciousness. We make the Multiprior tool freely available \cite{hirsch2019segmentation} in the hope that this will spawn further development and that it will be used in clinical research project.

\section{Methods}

\omitting{We use a 3D convolutional neural network (CNN) to capture detailed intensity pattern in the image -- the “Detail CNN”. It is enhanced with three types of spatial information: A tissue probability map (TPM) to capture the prior probability for each tissue at a given location; a “Context” network which processes a wide field-of-view (FOV) at low resolution; and a conditional random field (CRF) as post-processing step to implement known morphological constraints. }

\begin{figure*}[ht]
  \centering
  \includegraphics[width=0.9\linewidth]
  {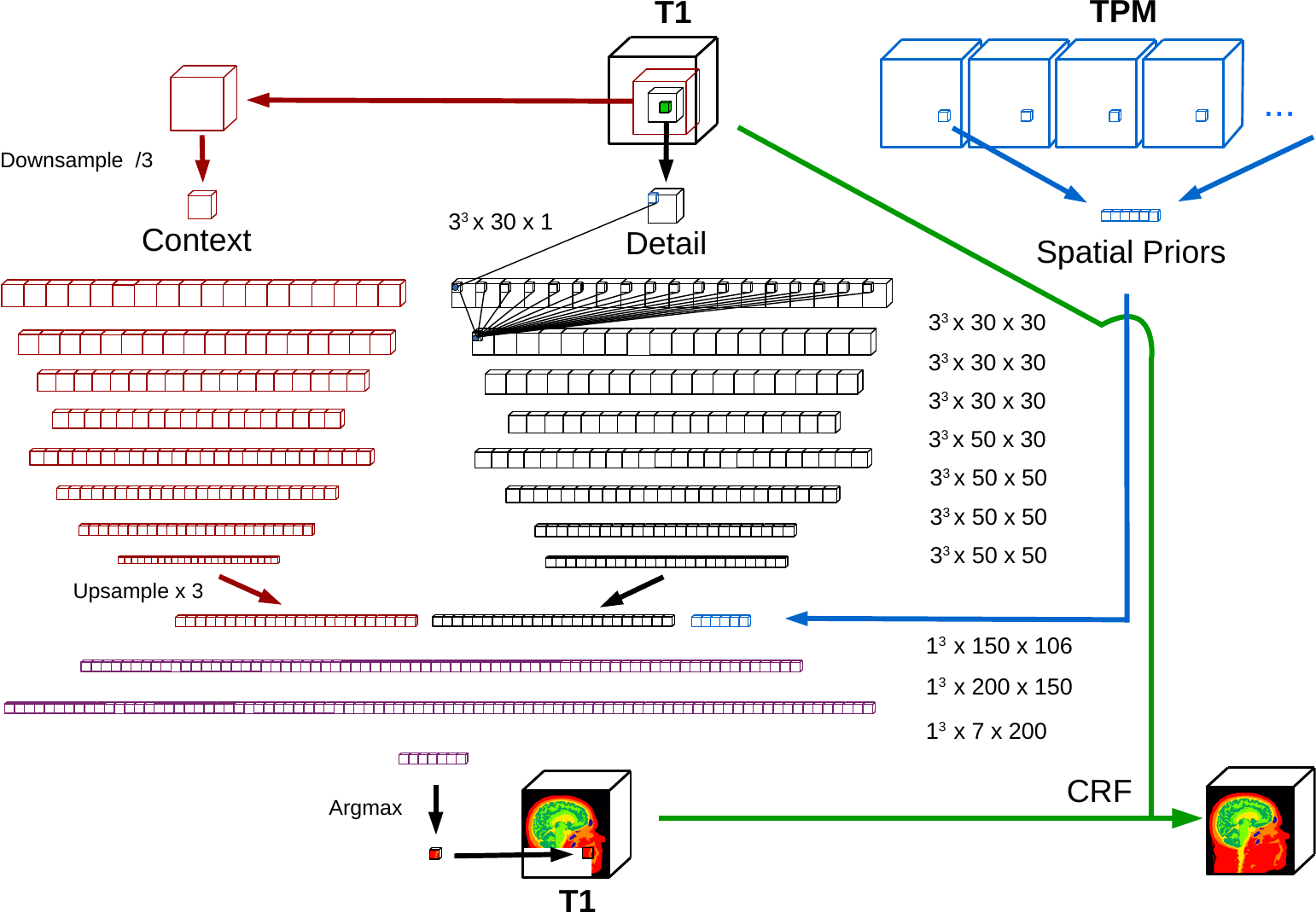}
  \caption{Multiprior network structure. The Detail network (black path) consists of a 3D convolutional neural network (CNN) with 8 layers. During training, this network takes as input a patch of $25^3$ voxels around a target patch of $9^3$ to be classified (green cube).  The size of the convolutional kernels mapping between layer is indicated by numbers to the right. For instance, 33x50x30 indicates a 3D convolution kernel of size $3^3$ transforming 30 features to 50 features. The “Context” network (red path) is identical in structure to the “Detail” network, except that it processes a downsampled version of a larger field-of-view of $57^3$ voxels during training. It includes an upsampling layer at the end in order to merge features at the same scale as the Detail network. Prior probabilities for the target patch are extracted from a tissue probability map (TPM) and added as input to the final classification (Blue arrow).  The “Classification” network (purple) takes the concatenated output of all three pathways as input and classifies the target patch with three fully-connected layers and no additional spatial mixing (kernel of size $1^3$). After the entire image has been segmented, a 3D conditional random field (CRF) processes the resulting output segmentation while taking the original input image into account (green arrow). Arrows indicate copying. }
  \label{fig:Multipriors}
\end{figure*}

\subsection{Detail CNN}
The Detail CNN (Figure~\ref{fig:Multipriors}, black path), aims to extract detailed intensity features at full image resolution. It takes as input a 3D image patch surrounding the target volume (green cube). The network consists of successive convolutional layers with kernels of size $3^3$. By limiting the convolutions to the patch boundaries the output of each layer reduces the patch size by 2 voxels in each direction. At the same time we increase the number of features extracted from this patch.  After 8 layers, an input patch of $17^3$ is reduced to an output of a single voxel -- this layer is said to have a ‘receptive field’ or $17^3$. Meanwhile, we increased the number of features to 50 per voxel. This spatial shrinking and feature expansion is a common practice for CNNs to convert spatial information into increasingly diverse and complex features. Algebraically, this network can be described as:
\begin{equation}
{\bf y}_i^{l+1}=f(\sum_j {\bf w}^l_{ij} * {\bf y}_j^l+b_i^l) \, ,
\end{equation}
where $*$ represents a 3D convolution, ${\bf y}_i^l$ represents the activity in a 3D patch (tensor) for feature $i$ in layer $l$, ${\bf w}_{ij}^l$  is a 3D convolution kernel connecting features $j$ from layer $l$ to feature $i$ in layer $l+1$, $b_i^l$ is a bias term, and $f()$  is a LeakyReLU as element-wise non-linearity. For the first layer, ${\bf y}_j^l$  represents the input to the network. For the last layer ${\bf y}_j^l$  represents the output (here $l=8$). Learnable parameters are the convolution kernels ${\bf w}_{ij}^l$   and bias terms $b_i^l$. These are adjusted using gradients of the cost function (see below). The size of the kernels ${\bf w}_{ij}^l$  is indicated as numbers between each layer in Figure~\ref{fig:Multipriors}.


\subsection{Context CNN}
Following \cite{kamnitsas_efficient_2017}, we also include a CNN that operates in parallel on a  wider FOV. The increased FOV allows the segmentation to rely on the surrounding context. For instance, in T1-weighted images of the head, areas with uniform black patches are not background, but instead \edit{ventricles} if they are in the interior of the skull. This wider FOV is first downsampled (by a factor 3). The network itself has then the identical structure to the Detail CNN. The output is then upsampled (by the same factor of 3). The size of the receptive field for this network including the downsampling is $51^3$. 

\subsection{Classification network}
Segmentation implies classifying each voxel into one of several possible tissue types. Here this is skin, skull, CSF, white matter, gray matter, air cavity, and background air - seven classes in total. The output of the preceding paths is concatenated and serves as the input to a final Classification network (Figure~\ref{fig:Multipriors}, purple). The Classification network has three  fully-connected layers with no spatial mixing. Algebraically the classification network can be described as:
\begin{equation}
{\bf y}_i^{l+1}=f(\sum_j  w^l_{ij}   {\bf y}_j^l+b_i^l) \, ,
\end{equation}
which is the same as the network defined in Equation (1) except we are no longer implementing convolutions on 3D patches, so $w^l_{ij}$ is no longer a 3D tensor. The index $l$ goes now from 9 (the output of the previous network)  to 11 (the classification output). We use again LeakyReLU as element-wise non-linearity  $f()$, except for the last layer, which uses a softmax function to encode the class probability. This output is used to compute the cost function during training, namely, the generalized Dice score in Equation (3) below. The final segmentation is obtained as the argmax operation of these seven tissue probabilities at each voxel.

\subsection{Tissue probability Map (TPM)}
\begin{figure}[h!]
  \centering
  \includegraphics[width=0.7\linewidth]{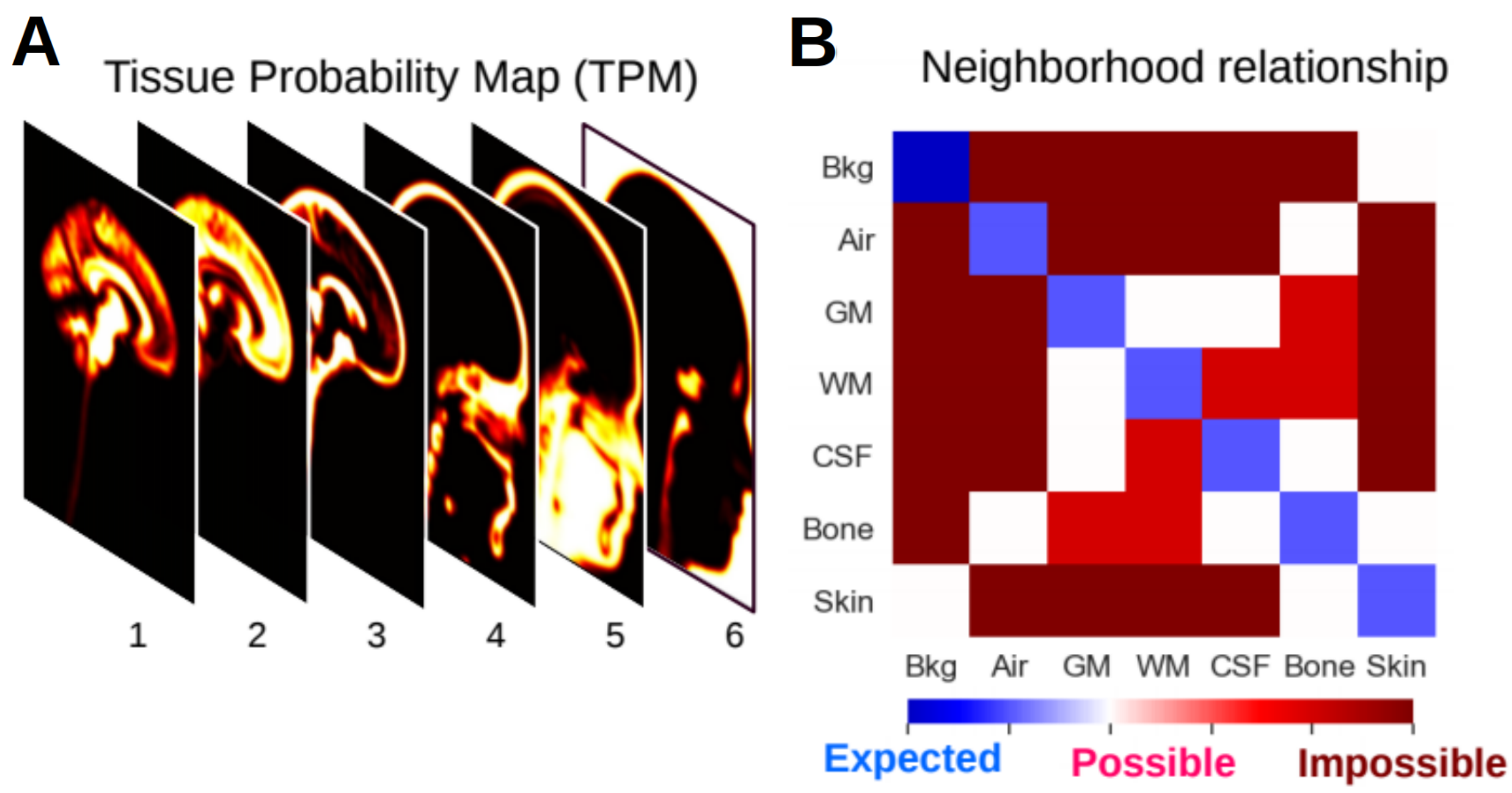}
  \caption{Location and neighborhood priors: (A) Tissue probability map (TPM) representing prior probabilities for finding each of 6 tissues at various locations: Gray Matter (1), White Matter (2), CSF (3), Bone (4), Skin (5) and background (6). (B) Penalty terms for conditional random field (CRF). Weights on this matrix represents the prior belief we have on observing a neighboring tissues. High penalties are specified for tissue pairs that are not supposed to be next to one another (impossible boundaries), intermediate penalties for tissues that are expected to be adjacent (possible boundaries), and low penalty for neighboring voxels to be in the same tissue class (expected continuity). Note that the TPM available to us has 6 tissue types,  while the CRF has 7, as we expect the classifier to distinguish between internal and external air, i.e. air cavities vs exterior background. }
  \label{fig:TPM}
\end{figure}

Spatial priors are included by providing the TPM values corresponding to the target voxel location (Figure~\ref{fig:Multipriors}, blue path). This requires registering the TPM to the individual MRI to find the corresponding locations. This was done with the warped coregistration of the segmentation routine in SPM8 \cite{ashburner_unified_2005}. We used a TPM that covers the full head down to the neck \cite{huang_automated_2013} as shown in Figure~\ref{fig:TPM}A. These prior probability values are given as input to the Classification network for each voxel as a vector with one values per tissue class. These values are concatenated to the output of the Detail network and Context networks. \omitting{The TPM we use here has prior probabilities for 6 types of tissues, yet we generate 7 classes (distinguishing air inside the head and outside air).}    

\subsection{Fully connected CRF}
Topological prior expectations such as connectivity of tissues are useful for producing realistic segmentation maps. A standard approach to implementing such priors is to use Conditional Random Fields \cite{kamnitsas_efficient_2017, chen_deeplab:_2018, schwing_fully_2015, chandra_dense_2017}. A CRF can assign an additive penalty to certain combination of segmentation labels.  In brain images we expected a-priori that tissues share a boundary with other tissues while others may not touch, e.g. white matter and gray matte or skin and bone are commonly found together, while we never expect to find cerebrospinal fluid next to air, or brain tissue next to skin. We express these prior probabilities of label configuration with a penalty matrix that is specific to each tissue pair (Figure~\ref{fig:TPM}B). This is more general than the conventional Potts model  commonly used with CRFs, and is similar to what we used previously for head segmentation \cite{huang_fully_2015}. The Potts model does not distinguish between class labels, and is typically used only for smoothing segmentation boundaries and removing small disconnected regions. By fully exploiting the flexibility of the penalty matrix we implement tissue-specific constraints. Here we use a fully-connected CRF, which allows segmentations to exert an influence on distant voxels \cite{krahenbuhl_efficient_2011}. Additionally, we extend to 3D the existing 2D implementations of fully-connected CRF \cite{krahenbuhl_efficient_2011}. The algorithm is initialized with the softmax output segmentation of the classifier and iterates to force segmentation labels to comply with the neighborhood constraints, while taking the single-voxel intensity of the original image into account. We iterate the CRF for 5 cycles. 

\edit{
\subsection{UNet}
We also implemented a volumetric version of the UNet \cite{ronneberger_u-net:_2015}, that takes as input image patches of size $32^3$ voxels, and learns volumetric features through 3 downsampling and subsequently 3 upsampling convolution blocks with residual connections. Each convolution block consists of two consecutive convolutional layers with filters of size $3^3$ and one downsampling or upsampling layer respectively. The TPM module was added before the last classification layer of the network, by simply concatenating the feature maps with the probabilities per tissue.
}

\subsection{Cost function}
Training was set to reduce the Generalized Dice Loss between the predicted segmentation by the network and the ground-truth provided by the manual segmentations \cite{crum_generalized_2006, sudre_generalised_2017}. The Generalized Dice Loss extends the conventional binary Dice Loss to the case of multiple target classes, and is defined here as:
\begin{equation}
D = 1 - \frac{1}{C} \sum_{i=1}^C \frac{2  {\bf y}_i \cdot  {\bf t}_i }{  {\bf y}_i \cdot {\bf y}_i+ {\bf t}_i \cdot {\bf t}_i} \, ,
\end{equation}
where \edit{${\bf y}_i$} is the output of the classification network (at \edit{the last} layer, $l=11$) and \edit{${\bf t}_i$} is the desired output classification (truth label in the training data with one-hot encoding), therefore $i$ now represents the class labels. The inner product $\cdot$ sums over all elements of the 3D volume. $C$ is the total number of classes ($C=7$ in this case). This loss function (3) is equal to zero when the network output matches the target classes in all voxels, and equal to one when no class probability is the same as the target in any location.  The generalized Dice loss intends to account for class imbalance often observed in image segmentations \cite{sudre_generalised_2017}. Our definition differs slightly in that the denominator takes the L2 norm of the class probabilities. This emphasizes large probability values and discounts small probabilities present in the vast majority of voxels, which tend to dominate in an L1 norm as used by \cite{sudre_generalised_2017}. \edit{We also omit class-specific weights here as they were not necessary for our data.} When we report performance we give only the second term (without the minus sign), or report individual terms of the sum when reporting the results for each class, which is the conventional binary Dice score for each class. 

\subsection{Training and testing data}
The training data consists of T1-weighted MRI scans from 4 healthy subjects and 43 individuals who suffered a stroke. The strokes occurred at least 6 month prior to the MRI scan, at which point the lesion is largely replaced by CSF. MRI scans from normal subjects were obtained on a 3T Siemens Trio scanner (Erlangen, Germany) (Huang et al. 2013). The stroke scans were collected at Georgetown University and the University of North Carolina, Chapel Hill, also on a 3T Siemens Trio scanner. The trained network was also applied to MRI images of 47 patients with disorders of consciousness collected at the Pitié-Salpêtrière University Hospital in Paris, on a 3T General Electric Signa system (Milwaukee, WI) \cite{hermann_combined_2019}. Image resolution was 1 mm in all three axis. 

Target labels for training and testing consist of semi-manual segmentations. Specifically, the 43 stroke heads are first segmented automatically \cite{huang_automated_2013, huang_realistic_2019} and then manually corrected for errors in particular around the stroke lesions and boundaries between CSF, gray matter and skull, resulting in 7 classes (background, air cavities, skin, bone, cerebrospinal fluid, white matter and gray matter). The T1 images from the 4 healthy subjects were segmented following the same procedure and have been previously published \cite{huang_automated_2013}. We used the graphic user interface of Simpleware ScanIP (Synopsys, Mountain View, CA) for manually correcting the segmentation. We obtained an additional independent manual correction for six of these stroke heads, starting from the same SPM8 results. Manual corrections in 3D were performed by technical staff at Soterix Medical Inc. as part of a clinical research project. \edit{Manual segmentations in 2D (see Section~\ref{sec:method-reader}) were performed in Simpleware from scratch by research staff at CCNY. All segmenters were trained on several images by author YH, based on \cite{ettarh_pocket_2014} and were supervised by YH during segmenting. }

\responseedit{During network training, 4 heads were kept out for validation purposes, measuring generalization performance during training epochs and used to define the stopping point, i.e., the epoch with maximum Dice score on the validation set. This procedure was used for training all convolutional neural networks (Multiprior, DeepMedic and U-Net variants).}  

\edit{
\subsection{Evaluation on an independent test set and reader study}
\label{sec:method-reader}
To evaluate the performance of the Multiprior network, we compared the segmentation results with traditional and state-of-the-art brain segmentation methods: FSL, SPM, and DeepMedic. We manually segmented 2D slices on an independent test set of 10 stroke heads and 10 patients with disorders of consciousness. For each head one slice was selected to include all tissue types and the major abnormal anatomy. \omitting{Note that QuickNAT only outputs segmentation for the brain tissues. Therefore there are no evaluation results for QuickNAT in the non-brain tissues. Also, segmentations are saved in the FreeSurfer standard space, so we registered and resampled that back into the MRI voxel space to be consistent with the other approaches.} To assess reliability of the human segmenters we obtained in addition a second independent human segmentation on 9 of these 10 stroke heads. 
}

\subsection{Statistics}
Performance of a model is evaluated on Dice scores of each scan, averaged over classes unless \edit{otherwise} noted. To evaluate statistical significance between results of two models, we compare the distribution of the Dice scores \edit{across} scans. Whenever the scan samples are equal we use the Wilcoxon rank-sum test. For unpaired data we use the Mann-Whitney U test. For multiple group comparison with paired data we use the non-parametric equivalent of repeated-measurements ANOVA, namely the Friedman chi-square test. Bonferroni correction was used to account for multiple comparisons when doing pairwise comparison of multiple models. \responseedit{A paired t-test was performed to compare Dice scores for the disagreement between machine and segmenters and disagreement between segmenters.}

\section{Results}

\subsection{Prior probabilities, morphological priors and spatial context all improve performance}
We designed a three-dimensional convolutional neural network for segmentation with the architecture defined in Figure~\ref{fig:Multipriors}. In its simplest form the network extracts detailed spatial features in a small volume around the voxel to be segmented  (Figure~\ref{fig:Multipriors}, Detail). We refer to this as the Detail CNN. The features extracted by this network are then classified by the Classification network into one of seven tissue types. We will consider three sources of prior information to be added to this classification network: image context by adding a low-resolution wide field-of-view input (Context, Figure~\ref{fig:Multipriors}); spatial prior information through a tissue probability map (TPM, Figure~\ref{fig:Multipriors}~\&~\ref{fig:TPM}); morphological constraints implemented with conditional random fields (CRF,  Figure~\ref{fig:Multipriors}~\&~\ref{fig:TPM}).

\begin{figure}[ht]
  \centering
  \includegraphics[width=0.6\linewidth]{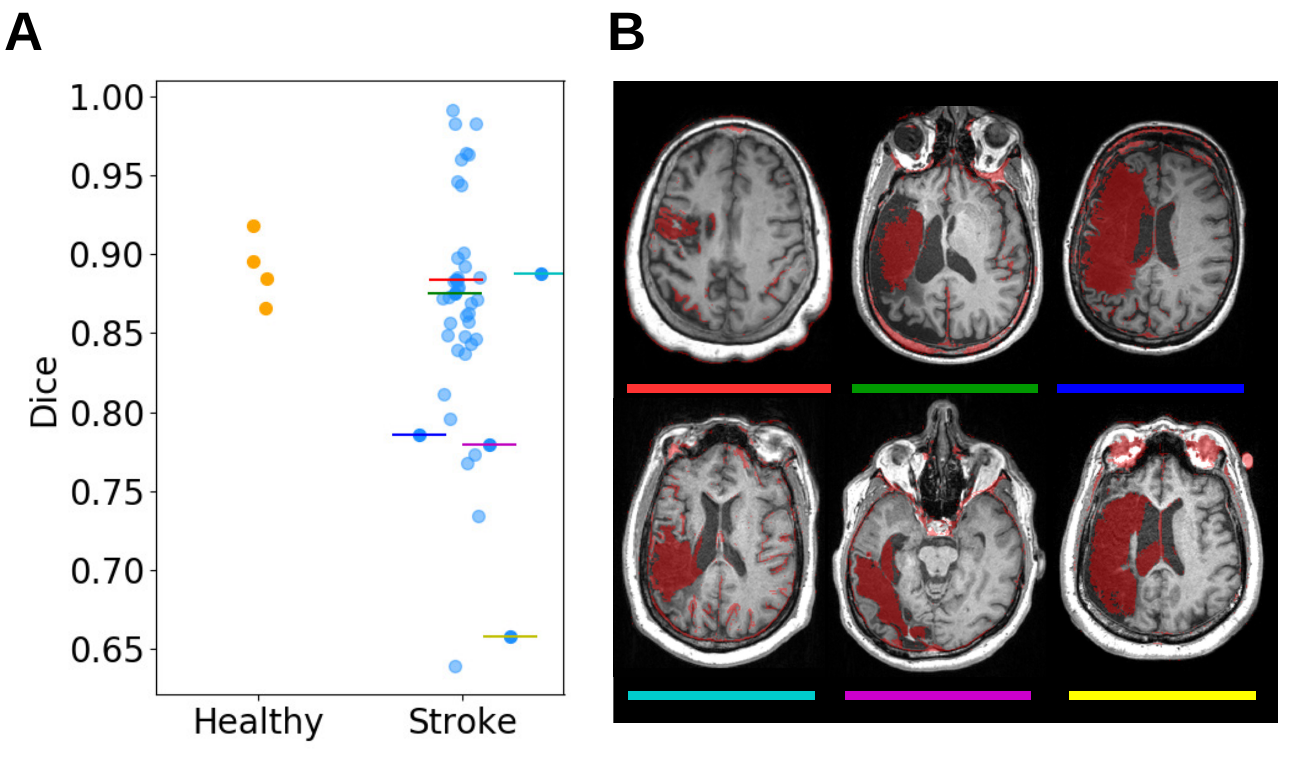}
  \caption{Semi-manual segmentation of T1-weighted MRI images of the head. (A) Dice score comparing SPM8 automated segmentation with segmentation after manual correction by an expert human segmenter. \responseedit{Each point represents the average Dice score across all 7 tissues (gray matter, white matter, CSF, bone, skin, air and background) in each subject.} Dice scores here can be seen as a performance metric for SPM8, or alternatively, as an indication of the level of manual effort of correcting for each volume. Generally only minor corrections were performed for the healthy subjects (N=4) as compared to some of the stroke patients (N= 43). Colored lines indicate 6 patients shown in more detail in panel B.  (B) Corrections of expert human segmenters (red) overlaid on T1-weighted MRI for 6 exemplary stroke cases. }
  \label{fig:Data1}
\end{figure}

Network parameters are trained on semi-manual segmentations from 43 stroke patients with chronic lesions and 4 healthy subjects with normal anatomy (data collected at different clinical sites; see Methods). These heads have been segmented using SPM8 and manually corrected by expert segmenters. Dice scores between the SPM8 results and the corrected segmentations give a sense of extent of the corrections (Figure~\ref{fig:Data1}A). They are minor in some instances, but substantial in others. Typically this manual correction process takes 16-32 hours of work for the entire volume (segmenting entire volumes by hand from scratch is prohibitive). Examples of the corrections are shown in Figure~\ref{fig:Data1}B. These focused on errors in the stroke lesions and gaps in thin structures that border on the resolution limit of 1mm in this data, e.g. skull, CSF, gray matter; \responseedit{all of which represent small areas relative to the whole head. The corrections are small on healthy subjects where SPM8 performs well, resulting in an average higher Dice score than stroke heads.} The \edit{manually corrected} segmentations were used here to train the network by minimizing the Generalize Dice Loss (Eq. 3) on training data (N=35 heads), and to evaluate network performance using the Dice Score on test data (N=12 head).\edit{ We repeat this on different train/test partitions of the data to obtain test set result for all 47 heads.} 

\begin{figure}[ht]
  \centering
  \includegraphics[width=0.6\linewidth]
  {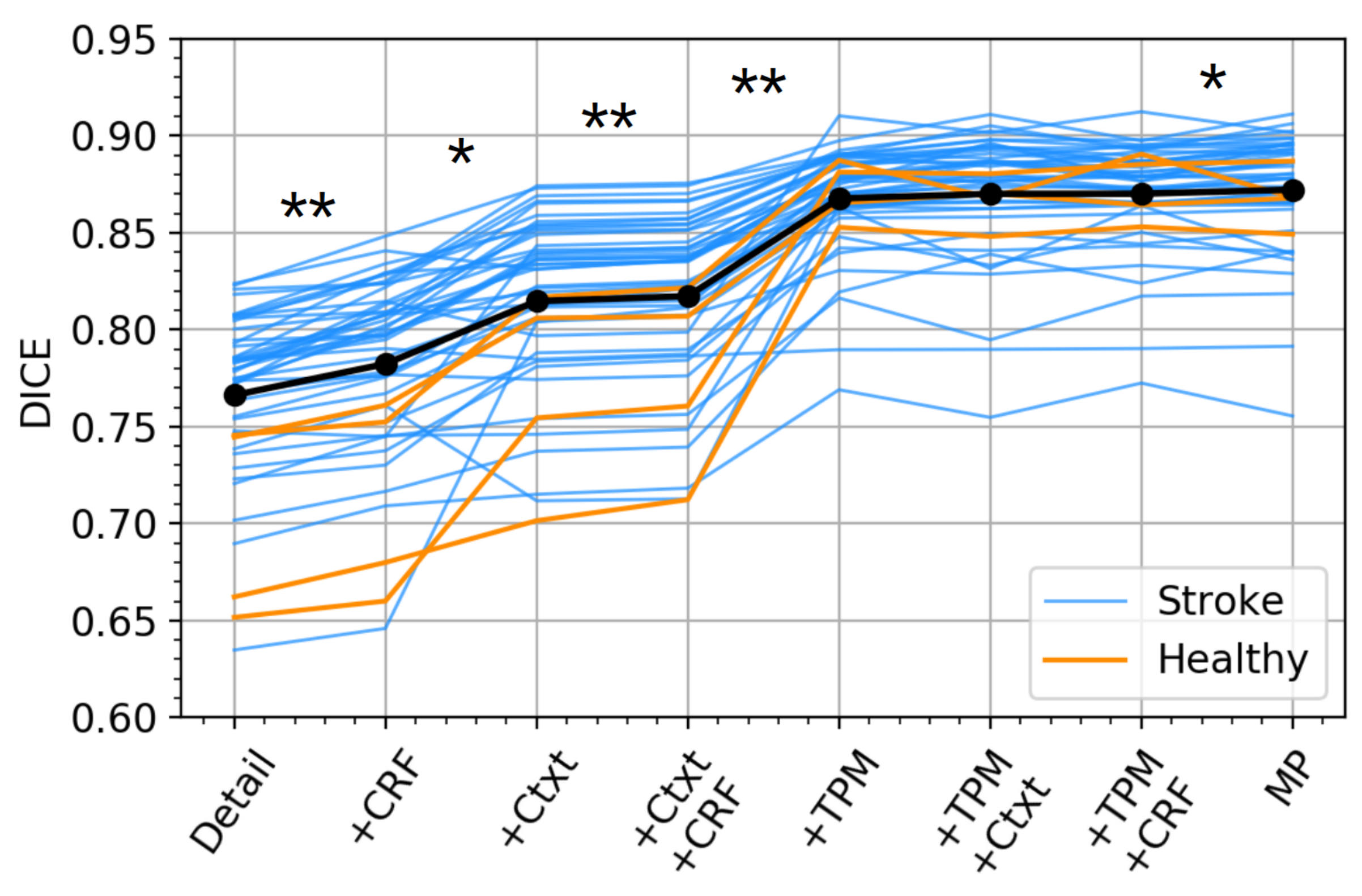}
  \caption{Additions of prior information or context improve segmentation performance.  Here we are showing the Dice Score average over all tissue types (second term of Eq. 3) for each test subject \edit{obtained with 4-fold cross validation on} 43 stroke and 4 healthy individuals \edit{(blue and orange lines respectively, and black line for the mean value)}. Dice score are measured between semi-manual and automated segmentations obtained with different network architectures. All networks contain the Detail CNN. Additions are a Conditional Random Field (+CRF), a Context CNN with wider field of view (+Ctxt), and a Tissue Probability Maps (+TPM). Multiprior (MP) is the combination of all networks (Detail+Ctxt+TPM+CRF). Stars indicate a statistically significant difference in Dice score between pairs of network architectures (* : p $<$ 0.05, ** : p $< 10^{-7}$, alpha level corrected for multiple comparison with Bonferroni correction and C=28 multiple comparisons).}
  \label{fig:AllModels}
\end{figure}

\edit{Test set performance is shown in Figure~\ref{fig:AllModels}.} 
To test for the benefits of each possible source of prior information we compare each addition individually to the results with the Detail network alone: Detail vs Detail+CRF (p=2.9 x $10^{-9}$, N=47, here and in the following all paired tests are rank-sum Wilcoxon test); Detail vs Detail+TPM (p=2.4 x $10^{-9}$); Detail vs Detail+Context (p=5.5 x $10^{-9}$). Evidently all sources of prior information significantly improve segmentation performance, with the largest numerical increase with the addition of the TPM (Figure~\ref{fig:AllModels}). Note that the Detail + Context network (denoted as +Context in \ref{fig:AllModels}) is essentially a previous approach known as the DeepMedic network \cite{kamnitsas_efficient_2017}. When combining all three enhancements (Detail + Context + TPM, + CRF), referred to here as the Multiprior network, we obtain the strongest performance. Notably, when removing the TPM from the Multiprior network performance drop significantly (p=2 x $10^{-9}$, N=47), as well as when removing the CRF (p=2 x $10^{-7}$), but not when removing Context (p=0.46). \edit{This analysis is the equivalent of an ``ablation'' study.}

\begin{figure}[ht]
  \centering
  \includegraphics[width=0.7\linewidth]
 {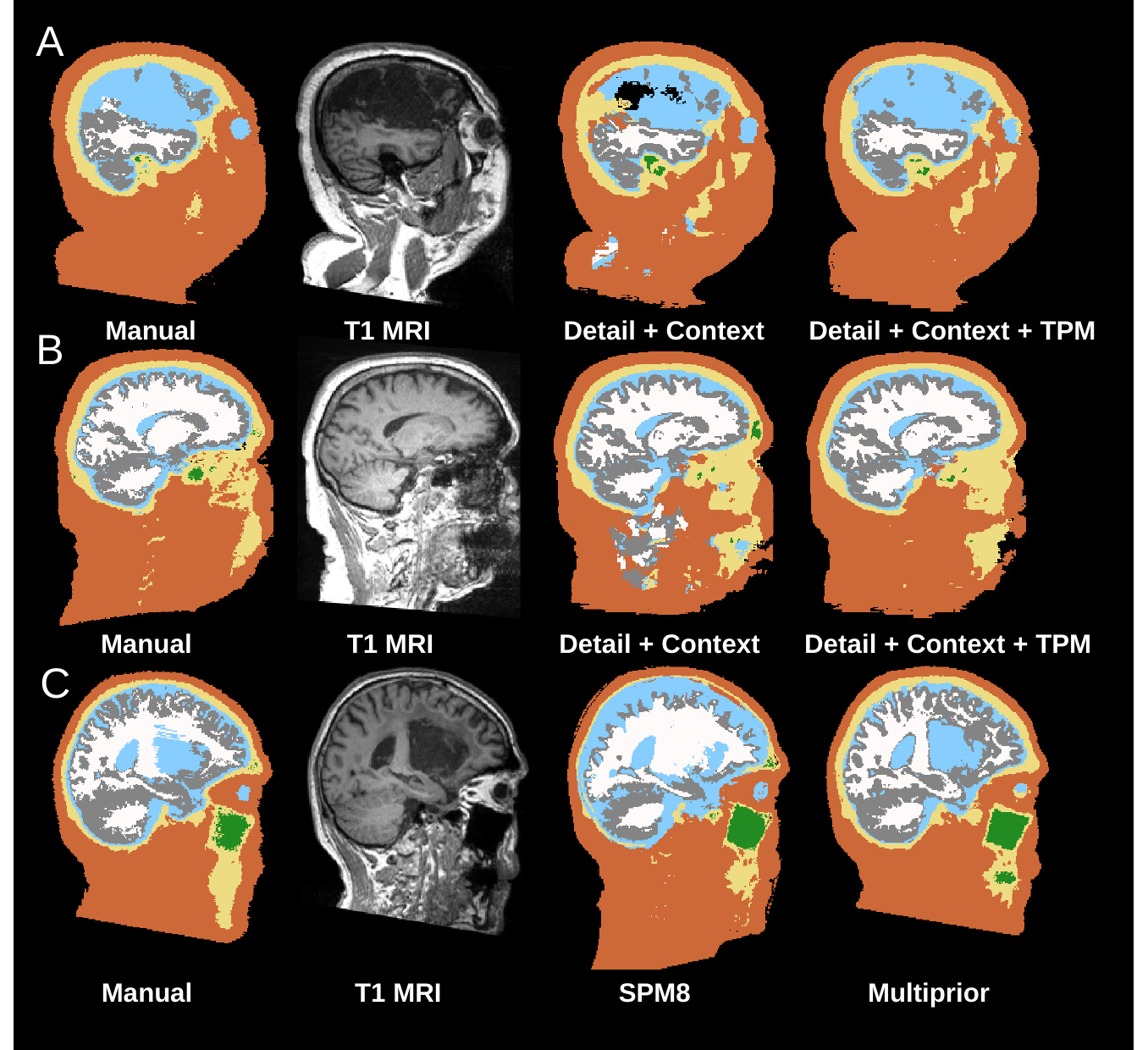}
  \caption{Examples of common segmentation errors for three stroke subjects and corrections by the Multiprior network. Each row corresponds to one subject. The manual segmentation is on the first column, followed by the T1 weighted MRI that is used as an input for the \edit{network}, next are segmentations from the Detail + Context network (A,B) and SPM8 (C) compared to the segmentation from the Multiprior. Each color represents one of each of seven tissue-classes used for classification: Black = background; brown = skin; yellow = bone/skull; green = air/sinus cavities; light blue = CSF; white = white matter and gray = gray matter. \edit{Notice the large CSF filled lesion in panel A.}}
  \label{fig:StrokeHeads}
\end{figure}

To gain a sense of the types of errors and corrections of each network we show sagittal cross-sections for 3 stroke cases (Figure~\ref{fig:StrokeHeads}). In the first example (Figure~\ref{fig:StrokeHeads}A) we see that the Detail+Context network err in labeling the CSF that has filled the stroke lesion as background air. Both are dark in the T1 weighted image, but the location inside the skull strictly prohibits this erroneous label, something that is readily corrected once adding the TPM.  The second example (Figure~\ref{fig:StrokeHeads}B) shows erroneous gray matter labels outside the skull or brain stem (in the neck), which again is readily corrected once the TPM is added. In the third example (Figure~\ref{fig:StrokeHeads}C) we show the overall benefits of the Multprior network as compared to the traditional segmentation with SPM8. Evidently SPM8 mistakenly labeled much of the dark CSF in the stroke lesion as white matter.

\begin{figure}[h!] 
 \centering
 \includegraphics[width=\linewidth]
 {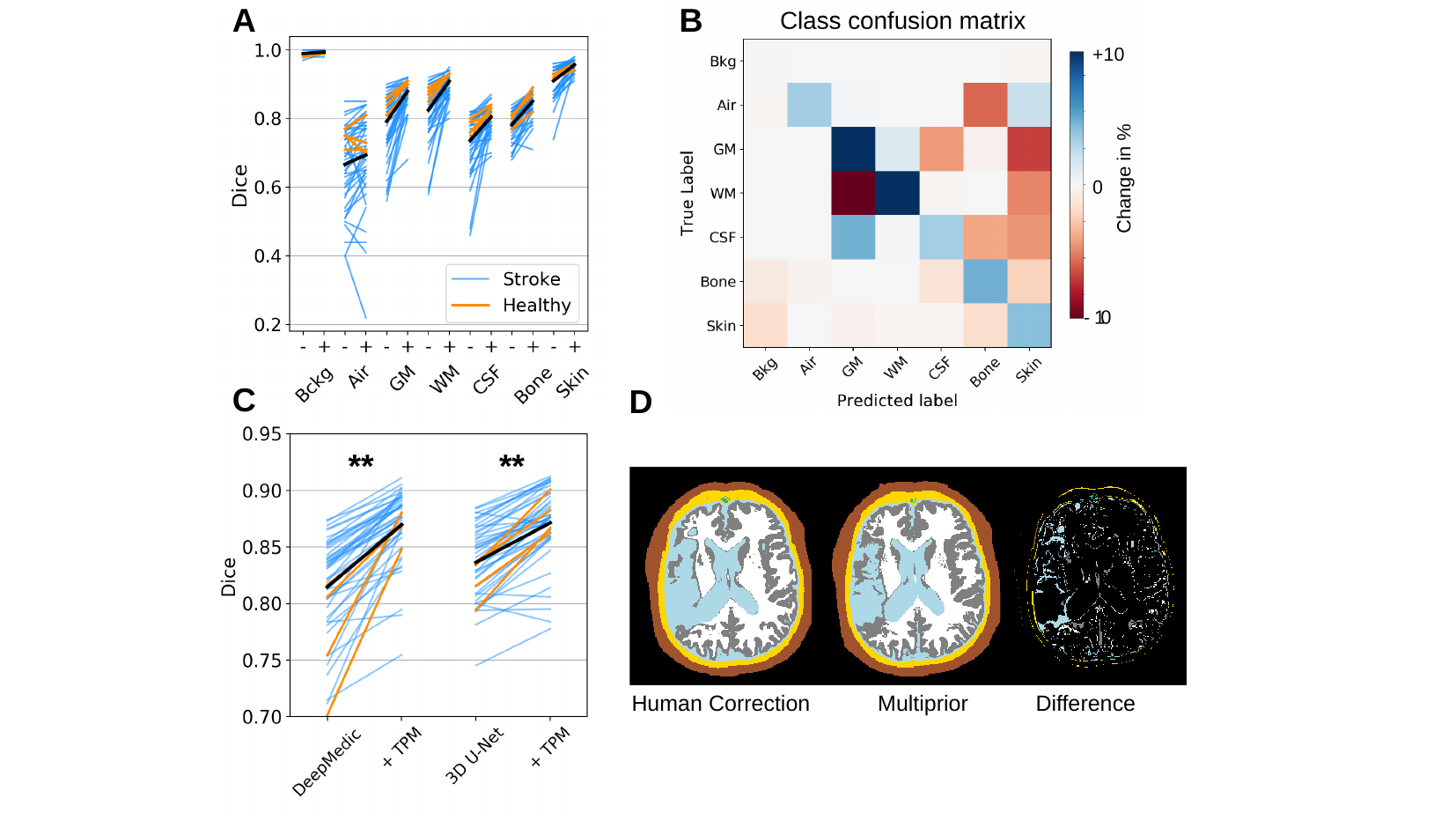}
  \caption{Benefits of \edit{adding a} Tissue Probability Map (TPM).  (A) Dice scores for each tissue class are shown for all 47 individuals for the Multiprior network including a TPM (+) or excluding the TPM (-). Black lines correspond to mean across all heads. Scans from healthy subjects are in orange \edit{and in blue for stroke cases.} (B) Change in the confusion matrix  when adding a TPM. Blue, positive values on the diagonal indicate more voxels correctly classified when the TPM is added; Red, negative values on the off-diagonal indicates fewer voxels that are incorrectly classified with the addition of the TPM. By definition, each row of this difference matrix must sum to zero, as corrections on the off-diagonal equate the improvements on the diagonal. (Bkg: background; Air: air cavities; GM: gray matter; WM: white matter; CSF: cerebrospinal fluid.)  \edit{(C) Benefit of adding a TPM for DeepMedic and UNet for the same stroke cases using Dice score for the entire head. DeepMedic is equivalent to Detail CNN + Context} (** : p $< 10^{-7}$). \responseedit{(D) Difference between manual segmentation and automatic segmentation with the MultiPrior network. The errors are mainly along boundaries, which can be extensive due to cortical folding, as well as the boundary of the stroke lesion with ambiguous intensities.} }
  \label{fig:AddingTPM}
\end{figure}

Since the addition of the TPM leads to the most significant improvement we analyze this result in more detail (Figure~\ref{fig:AddingTPM}). All tissue types improve substantially in performance (p $< 10^{-8}$; Figure~\ref{fig:AddingTPM}A), only for air in sinus cavities and background the results are somewhat mixed but still statistically significant (p $<$ 0.0003). When inspecting the changes in the confusion matrix (Figure~\ref{fig:AddingTPM}B), we see that the most significant corrections are in correctly distinguishing between white and gray matter, between air cavities and bone (both black in T1-weighted images), and generally correcting erroneous “skin” labels (a catch-all label for other soft-tissues) which is readily confused without prior knowledge.  

\edit{Adding a TPM should confer a generic benefit to other convolutional networks. To test this,  we adapted the popular UNet structure \cite{ronneberger_u-net:_2015} for the current task and added at the last classification layer an input from a TPM (see Methods). Adding a TPM to the UNet shows a significant increment in the performance on the stroke dataset (Figure~\ref{fig:AddingTPM}C; p = 3 x $10^{-9}$, Wilcoxon ranksum test). For reference we also show the summary for the whole head for the DeepMedic structure with and without an added TPM. As we saw previously this also results in a significant improvement (p = 7 x $10^{-9}$, Wilcoxon ranksum test). Both networks were trained with identical data and show comparable performance once the TPM is added. }

\subsection{Multiprior network \edit{matches human segmentation performance}}
\responseedit{Segmenting MRI images is a subjective process even for expert human segmenters \cite{egger_mri_2017, qiu_reproducibility_2019}} This leaves us without an \edit{absolute} ground truth to evaluate machine performance. It is therefore customary to take the agreement between independent manual segmentations as a point of reference. 

\begin{figure}[h!]
 \centering
 \includegraphics[width=\linewidth]
 {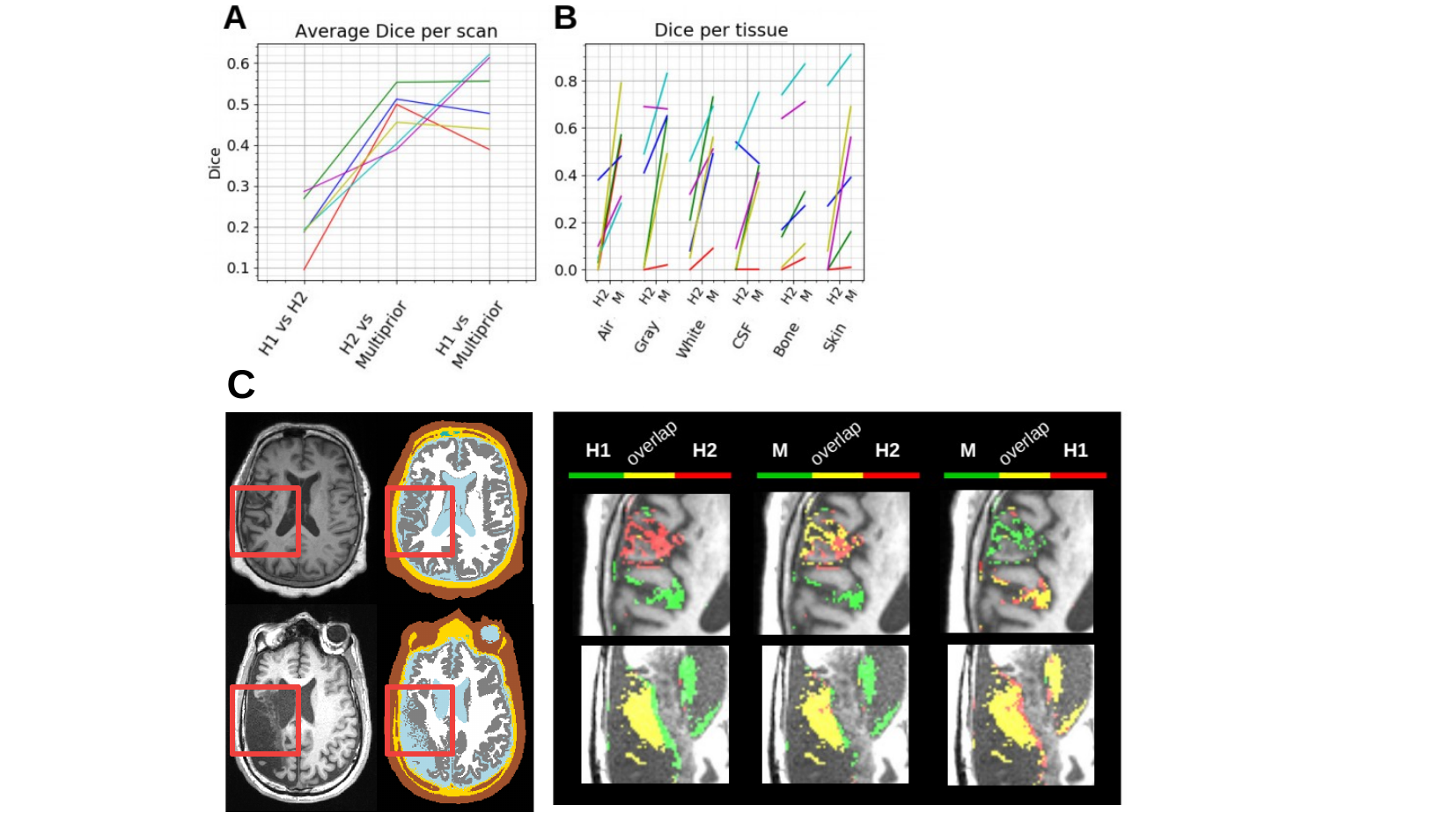}
  \caption{\edit{Multiprior network and human manual corrections in 3D volumes.} (A) Dice scores comparing two humans segmenters to one another (H1, H2) and to the Multiprior network.  For this test, six stroke heads (colored lines) were manually corrected independently by the two human segmenters.  The network was trained on a different set of 41 segmentations. (B) Dice scores between H1 and H2 or Multiprior network (M) now separated by tissue class. (C) \responseedit{Two example scans (first column), uncorrected segmentation from SPM8 (second column), and corrections of segmentation shown with} overlaps between Humans 1 and 2 (third column), Human 1 and Multiprior (fourth column) and Human 2 and Multiprior (fifth column). In the example of the first row the human segmenters chose to make corrections in different locations. In the second row corrections overlap (yellow) but were more extensive for H1. In both instances the machine segmentation overlaps more extensively with each of the human corrections as compared to the two humans (more yellow).}
  \label{fig:HumanPerformance-correction}
\end{figure}

\edit{For a full 3D evaluation we} obtained an additional independent manual correction for six of the stroke heads, starting from the same automated SPM8 results. We focus the evaluations on the areas that have been corrected by at least one of the two human segmenters (e.g. Figure~\ref{fig:Data1}). Dice scores between the two human segmenters in these areas are relatively low (Figure~\ref{fig:HumanPerformance-correction}A) and increase substantially between the human and the machine segmentation using the Multiprior network (p=0.011, Friedman-chi square, F=9.0, N=6).  This suggests that wherever the human decided to make a correction, the machine provided a similar segmentation, but the two humans did not make the same set of corrections. The result is similar when evaluating Dice score for individual tissues (Figure~\ref{fig:HumanPerformance-correction}B). An example of this for CSF is shown in Figure~\ref{fig:HumanPerformance-correction}C.  

\begin{figure}[ht]
  \centering
  \includegraphics[width=\linewidth]{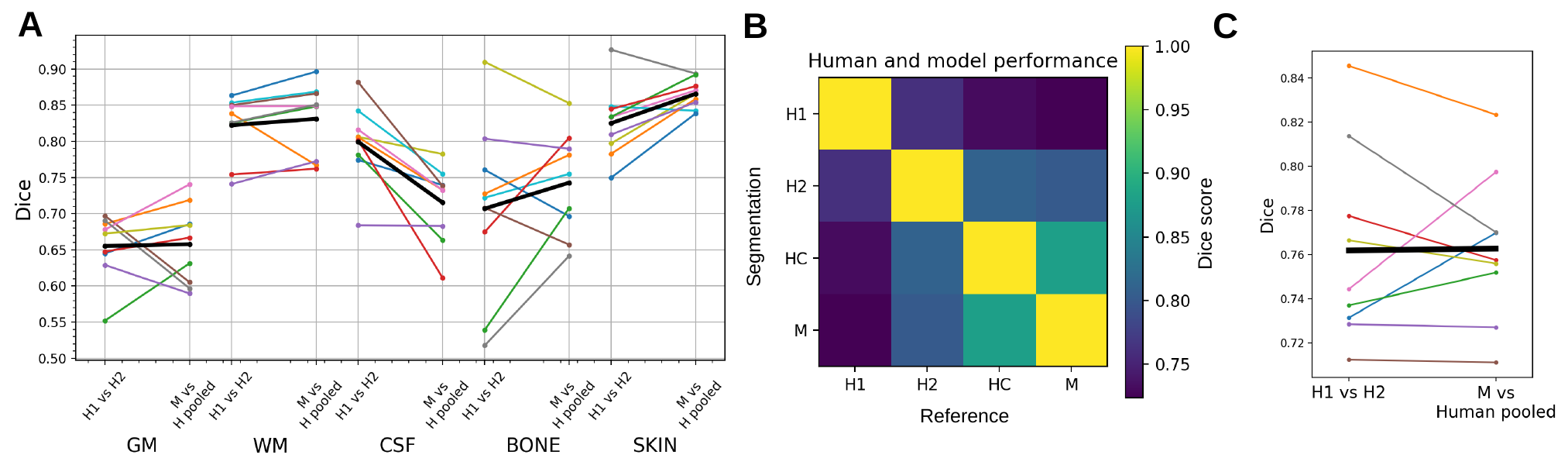}
  \caption{\responseedit{Multiprior network and human manual segmentation in 2D slices. (A) Dice score comparison for 9 stroke images that were not part of the training set, for five tissues (gray matter -GM-, white matter -WM-, cerebrospinal fluid -CSF-, bone and skin). Comparison is made between fully-manual segmentation (H1, H2) and MultiPrior (M) against the human segmentation (pooled average between H1 vs M and H2 vs M). (B) Dice score comparing fully-manual segmentations (H1, H2), manual correction of automated segmentation (HC) and Multiprior segmentation (M). The Dice score is averaged over all tissues and heads for all possible pairwise comparisons.  (C) Comparison of human fully-manual segmentations with Multiprior segmentation (for human-machine comparison Dice score are pooled over H1 and H2).}}
  \label{fig:HumanPerformance-scratch}
\end{figure}

The previous evaluation focuses on correcting an automated segmentation (from SPM8). To evaluate performance without the bias of the automated segmentation we also obtained \responseedit{fully}-manual segmentations. \responseedit{As this is a laborious process, and because manual segmentations are performed naturally one slice at a time, we restricted this evaluation to 2D segmentations}. Two individual manual segmentations were obtained for axial slices of 9 stroke heads that were not part of the training set. \responseedit{These were segmented by research staff at CCNY and were reviewed by authors YH and LH. Additionally, for these heads we had the semi-automated segmentations that were left out from the training set. Given that there is no absolute ground truth, we compare all segmentations to one another by computing the Dice score between all pairs including the Multiprior segmentations. We compare the Dice score of these comparisons for each tissue (Figure \ref{fig:HumanPerformance-scratch}A), and for all tissues combined (Figure \ref{fig:HumanPerformance-scratch}B). Overall, the disagreement between machine and segmenters is comparable to that of the disagreement between segmenters (see Figure \ref{fig:HumanPerformance-scratch}C) (mean of Dice H1 vs M and H2 vs M = 0.763, Dice H1 vs H2 = 0.762, paired t-test: t(8)=-0.08, p=0.94).
Evidently, even the human segmenters disagree in their judgement (see Figure 1 in the Supplement) with H2 being closer to the semi-automated and automated methods as compared to H1 (Fig. \ref{fig:HumanPerformance-scratch}B). Discrepancies between the manual segmentations are mainly due to ambiguities around lesions, where it is hard to decide where CSF ends and where gray matter starts, or how to characterize ``gray'' tissue in T1 images for areas where one would expect white-matter and which are likely pathological. Other discrepancies are on how to characterize bright cancellous bone (fatty soft-tissue is equally bright) and where the exact boundary is between sinus air cavities and cortical bone, or cortical bone and CSF, which are equally dark in T1 images. }

\subsection{\edit{Utility to a different clinical population of}  patients with disorders of consciousness}

\begin{figure}[ht]
 \centering
 \includegraphics[width=0.6\linewidth]
 {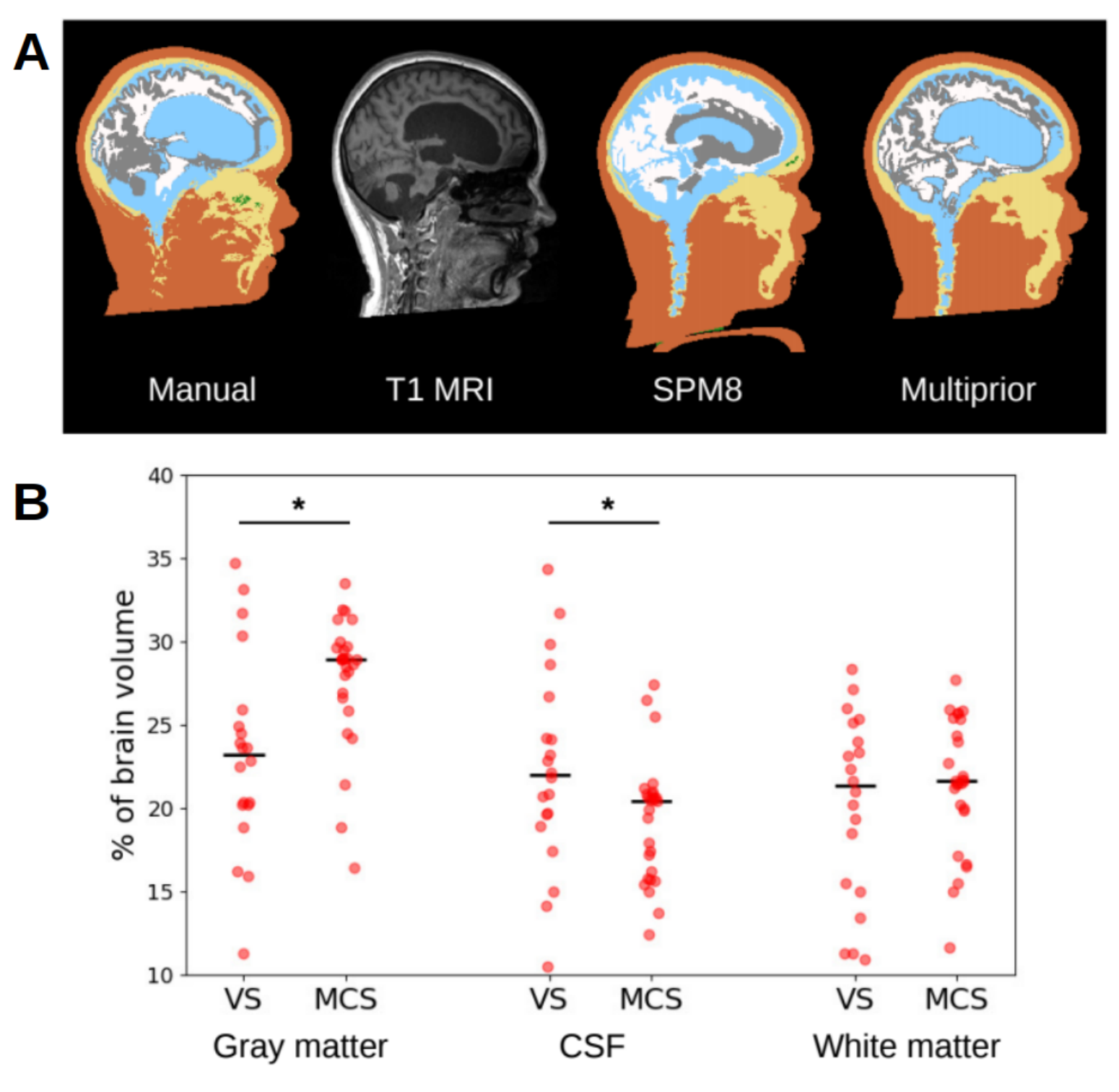}
  \caption{\edit{Segmentation of head and brain for patient in minimally conscious state.}   (A) Segmentation is based on T1-weighted MRI. The Multiprior network correctly identifies CSF (light blue) in the enlarged ventricles. SPM8 mistakes some of this as gray matter (gray) as it expects smaller ventricles typical for normal anatomy.  As a consequence this also affects recognition of white and gray matter elsewhere. (B) Anatomical differences between vegetative and minimally conscious state. Subjects in vegetative state (VS, N=20) differ from subjects in a minimally conscious state (MCS, N=27) on relative volume of gray matter tissue, cerebrospinal fluid (CSF), but not white matter volume ( * : p$<$ 0.05).}
  \label{fig:DOCPatients}
\end{figure}

To demonstrate the utility of the Multiprior network in practical use, we tested it on a \edit{different clinical population}. Specifically, 
we used the Multiprior network trained as before (stroke and healthy) to segment a new set of MRI images from patients with disorders of consciousness collected at a different clinical site. This included patients in vegetative state (VS, N=20) and minimally conscious state (MCS, N=27) \cite{hermann_combined_2019}. MCS patients show intermittent signs of consciousness and have higher chances of improvement. In contrast, VS patients exhibit no signs of consciousness and usually have a worse prognosis. A clinical diagnosis distinguishing between these two conditions is not trivial and some anatomical features have been shown to correlate with the patient’s \edit{diagnosis as VS or MCS} \cite{schnakers_diagnostic_2009, bender_persistent_2015}. The segmentation task is challenging because these patients have severe anatomical abnormalities (Figure~\ref{fig:DOCPatients}A). We segmented these heads with the trained Multiprior network, quantified the volume of different tissues and compare that to a clinical diagnosis of their level of consciousness. 

We find smaller volume of gray matter and larger volume of cerebrospinal fluid in vegetative state patients as compared to minimally conscious patients (Figure~\ref{fig:DOCPatients}B, N=47, p=2x$10^{-3}$ and p=2x$10^{-2}$, Mann-Whitney U test for unpaired data). Both these metrics have been previously linked to these conditions \cite{schnakers_diagnostic_2009, bender_persistent_2015}. In contrast, white matter volume, not an established biomarker, showed no significant difference (Figure~\ref{fig:DOCPatients}B, p=0.26, N=47). This demonstrates the utility of this automated segmentation approach in clinical applications.

\begin{figure}[h!] 
  \centering
  \includegraphics[width=\linewidth]{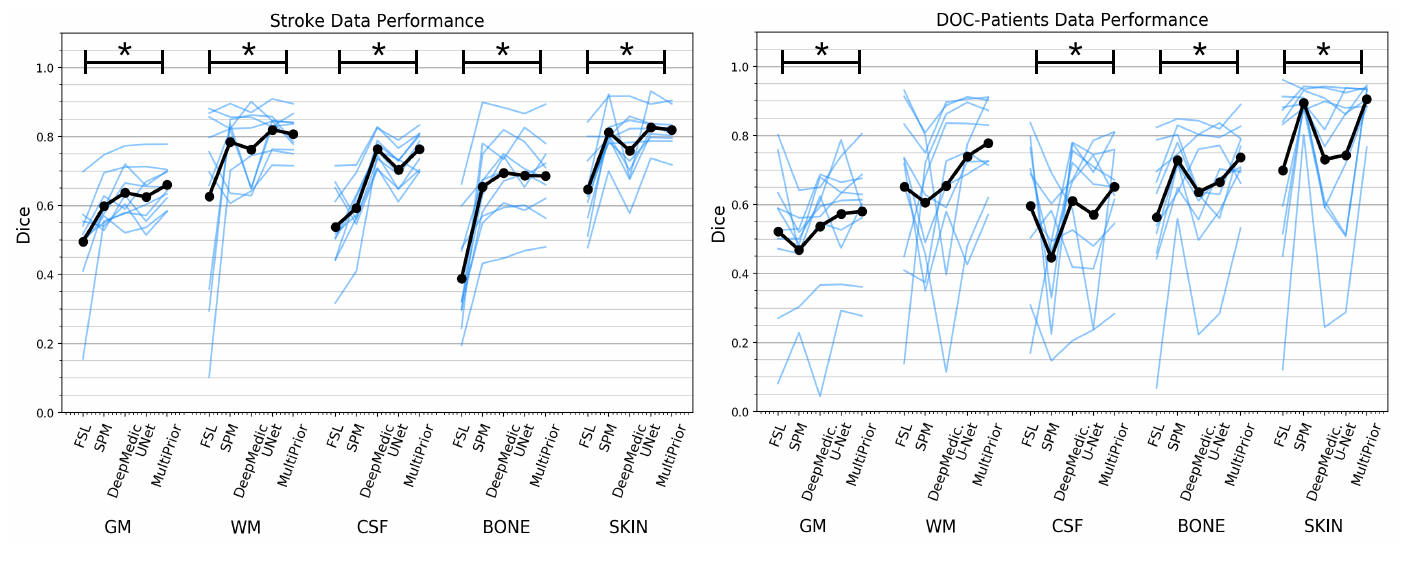}
  \caption{\edit{
Comparison of multiple segmentation methods against 2D manual segmentations. (Left) stroke patients (N=10) (Right) disorder of consciousness  patients (N=10). Methods are compared on the five different tissue classes (gray matter, GM, white matter, WM, cerebrospinal fluid, CSF, bone and skin). \omitting{ Because QuickNAT does not provide segmentations for non-brain tissue, these are not included in the figure.} Significant differences between Dice scores across methods is  indicated with * (p $<$ 0.05, repeated measurements ANOVA)}}
  \label{fig:Stroke_DOC_SPM_FSL_quickNAT_models_performance}
\end{figure}

\edit{
\subsection{Multiprior netwoork outperforms existing segmentation approaches on abnormal brain anatomies}
Finally, we wanted to compare the performance of the Multiprior model with existing automated segmentation methods, including traditional approaches (SPM and FSL), as well as newer deep convolutional networks (\omitting{QuickNAT and}DeepMedic \responseedit{and U-Net}). Given the present focus on lesioned anatomies we evaluated this on new manually segmented 2D slices for the stroke dataset (N=10) and the disorder of consciousness patients (N=10). Note that this also provided an out-of-sample evaluation, as the Multiprior network was trained on the stroke patients but tested here also on disorder of consciousness patients. Figure~\ref{fig:Stroke_DOC_SPM_FSL_quickNAT_models_performance} shows the result sorted by overall performance. In the average over all tissues, Multiprior numerically outperforms \responseedit{the UNet}, the DeepMedic followed by SPM and FSL (with mean Dice 0.73, \responseedit{0.73}, 0.72, 0.68, 0.54, respectively for the stroke dataset, and 0.73, \responseedit{0.66}, 0.63, 0.63, 0.61, respectively for the disorder of consciousness (DOC) dataset). \responseedit{The Multiprior network has similar performance to the UNet on the stroke data, and  performs better  numerically  than the UNet on the DOC patients.} On healthy tissues with normal anatomy all these tools perform well (on GM, WM, Bone, Skin). Lesions in this dataset are present as CSF-filled regions, which is where these tools fail to recognize the abnormality. Performance on CSF in stroke is the single instance where the DeepMedic numerically outperforms the Multiprior. \omitting{Note that DeepMedic and Multiprior were trained on the present stroke/healthy data, whereas QuickNAT was used in its pre-trained version  intended for normal anatomy \cite{ai-medquicknatv2_2020}. FSL and SPM have also been developed for normal anatomy, yet seem to be more robust in the presence of lesions than QuickNAT.} 
}

\section{Discussion}

The main contribution of this work is a network architecture that can segment the head anatomy with human-level performance \responseedit{specifically} in the presence of brain lesions. \responseedit{When tested on normal head anatomy of the proposed architecture, it is no worse than current state-of-the-art automatic segmentation software, and outperforms these methods in the presence of abnormal anatomy and lesions}, including an out-of-sample dataset. We provided the first detailed analysis of the relative merits of each type of spatial priors equivalent to an ``ablation'' study. In particular, we have shown that adding a prior probability map can benefit several existing network architectures. The same pre-trained network performed well on two different clinical populations, and we therefore make it freely available to facilitate broader testing.

Convolutional neural networks are by design shift invariant. They are thus able to recognize objects regardless of the position in an image. This has made these networks a powerful tool in parsing of images, such as those found in the Imagenet database \cite{imagenet_cvpr09}, where objects have no intrinsic correlation with their absolute position. This is however not the case in biomedical images, where anatomical structures are in a well-defined location relative to other tissues and organs. Prior knowledge of location is probably just as important as image intensity in clinical diagnosis.   

In T1-weighted MRIs, the CSF filled area in the brain created by a chronic stroke, appears black with a similar intensity to that of background and air-filled sinus cavities. Despite similar intensity, a human observer can readily discern the difference based on locations: black areas inside the head are either sinus cavities or CSF-filled lesions of the brain. Therefore, prior probability maps can easily resolve such conflicts. Indeed, we find that the largest gains can be achieved with a tissue probability map. \edit{This benefit of TPM was not specific to the network architecture we focused on here, which was based on the DeepMedic architecture. We found a comparable gain in performance when adding a TPM to a UNet segmentation architecture.} A convolutional network that does not take prior information into account should in theory still be capable of correcting these mistakes in anatomy, by enlarging the field-of-view to recognize relevant contextual information for each region. Therefore, learning this context information also improved segmentation performance here. 

The performance of the final models is in the range of \responseedit{75-92\% in Dice score (Figure 4)}. Most of the residual errors are edge/boundary mistakes, which dominate in brain images, due to cortical folding \responseedit{(Figure \ref{fig:AddingTPM}D), as well as errors in stroke lesions with ambiguous intensities and abnormal anatomy.} Labeling the highly folded surface of the cortex and gray-matter is a laborious and subjective  task, even for human segmenters, given that biological boundaries are seldom clear. For instance, in T1-weighted images bone and CSF are both dark and the boundary between the two must be surmised, rather than derived from the image intensities. An important criterion in this context is smoothness and continuity of these thin structures. Indeed, the addition of the conditional random field to the network serves this purpose and improved performance whenever it was added to the network.  

Aside of being subjective, manual segmentation is also very labor intensive. Segmenting an entire head \edit{volume} from scratch takes typically two to four weeks of labor\responseedit{, in particular if one wants to avoid typical continuity errors between slices as the manual segmentation is done one slice at a time.} The process we used here of only correcting automated segmentations takes about two to four days of work. It is therefore not surprising that two human segmenters chose to make corrections in different locations of the automated segmentations. The main advantage of the Multiprior network is that it has no such penalty for “effort” and segments the entire head from scratch. It matches the corrections of both human segmenters better than they match one another, as they often choose to make corrections in different locations. In this sense, the Multiprior \edit{improves on the corrections of human segmenters. When manual segmentations are performed from scratch on 2D slices the performance of the Multiprior network is equivalent to that of the human. The two human raters deviate in their segmentations from one another as much as they deviate from the segmentation of the network (Figure \ref{fig:HumanPerformance-scratch}).} \responseedit{However, a caveat of this study is that we only performed comparison on 2D slices instead of fully 3D segmentations, which are time-prohibitive.}

\edit{
Comparison of common segmentation tools with Multiprior network was done also on 2D manual segmentations of two different patients populations. On these data normal tissue is segmented similarly by all methods, while lesions containing CSF is only recognized by our deep convolutional networks. Both DeepMedic and the Multiprior surpass SPM and FSL \omitting{ and QuickNAT} (Figure \ref{fig:Stroke_DOC_SPM_FSL_quickNAT_models_performance}). \omitting{Note that the QuickNAT model we used here was trained on heads without any lesions, which explains its lower performance.} }


In this work we have extended the conditional random field \cite{krahenbuhl_efficient_2011} to work over four dimensions (intensity and three spatial coordinates). We also added multiclass conditional probabilities, which are particularly well suited to implement anatomical constraints. The CRF was added as a post-processing step. Recent work recasts the CRF computations into a recurrent convolutional neural network \cite{zheng_conditional_2015}. While we have not implemented this here, a fixed number of iterations of this recurrence can be readily added as an additional layer to the Multiprior network. In this way the fully-integrated network could be trained coupling the learning of the parameters of the fully-connected CRF with the learning of the parameters of the deep convolutional network. This would allow to train the spatial priors (unary potentials in the CRF) as well as the neighborhood priors (pairwise potentials), along with the features extracted.

\responseedit{We note that the biggest improvement was achieved by adding the location priors in the form of a TPM (Figure 4). \omitting{However, adding the other two priors (CRF+Context) can still significantly improve the performance after adding the TPM. This suggests that each module aided in correcting specific and different errors, and that no module can be left out.} In this dataset, the deep network improved most with the tissue-specific location priors. It is possible that on a different dataset, a TPM is less beneficial than the other two sources of prior information, but this remains to be tested in future work.}

\responseedit{One caveat to the proposed approach is that the tissue probability map has to be spatially aligned to the images before segmentation. In contrast, neural-network methods typically do not require this initial pre-processing and can operate in a stand-alone fashion. We have used SPM8 for this initial co-registration to warp the TPM to the target image, but we could have used other powerful warped registration methods such as ANTS \cite{avants_symmetric_2008}. Future work may incorporate newer network methods that implement registration as part of the network architecture \cite{jaderberg_spatial_2015}. We note that incorporating the CRF and co-registration into the network itself adds significant algorithmic complexity. Avoiding this added complexity was a deliberate choice given that they can be readily implemented as pre- and post-processing steps.} 

\omitting{Traditional generative models are well suited to combine prior information. For instance, the algorithm used in SPM takes prior location information into account \cite{ashburner_unified_2005} and a CRF incorporates morphological prior information \cite{kamnitsas_efficient_2017}. Generative models can also readily combine prior information on location with morphological information  \cite{huang_fully_2015} or shape information \cite{tu2008brain}. The current work is the first to combine prior information in a deep network model in the context of MRI head segmentation.}

In general, as the amount of training data increases, there is decreased need to implement prior information. We currently specified prior information “by hand”, namely, neighborhood priors were selected based on anatomical knowledge, and location priors were taken from existing tissue probability maps. However, the network architecture we have developed here can in principle also update these priors based on additional training data. As such, one can conceive of the Multiprior architecture as having dedicated storage modules to memorize prior information in a concise and interpretable code. As long as datasets are limited in size, explicit priors will aid segmentation performance. And even if data volumes grow in time, good initial priors will remain useful starting points to accelerate training. 

All source code \edit{and pre-trained networks are} available at \url{github.com/lkshrsch/MultiPrior_Brain}.

\section*{Disclosures}
No conflicts of interest, financial or otherwise, are declared by the authors.
The present study was exempt from IRB review at the City College of New York as it only used preexisting de-identified data.

\section*{Acknowledgment}
The authors would like to thank Chris Thomas at Soterix Medical Inc. for manually segmenting most of the 41 stroke heads and providing the data. We would also like to thank Peter Turkeltaub, Adam Woods, and Abhisheck Datta for collecting and sharing the stroke MRI data. Similarly we would like to thank Jacobo Sitt and Bertrand Herman to sharing the MRI data from the minimally conscious patients.  \edit{We also want to thank Noor-E-Jannat Anindita and Humna Khan from the City College of New York for manually segmenting 2D slices of 20 heads.}

\bibliographystyle{IEEEtran_max_6}
\bibliography{IEEEabrv,Multipriors}

\newpage

\appendix

\section{Examples of manual segmentations}

Volumetric segmentation labels were obtained in a semi-automatic process using SPM8 and applying manual corrections, which were used to train the neural network. In order to have an unbiased set of labels; manual segmentations were done from scratch in 2D slices containing visible lesions. To obtain a reference for variability between manual segmentations, ten scans were segmented twice, each independently by a separate segmenter. We show five examples of these duplicated segmentations, displaying regions were there is disagreement. 

\begin{figure}[h] 
  \centering
  \includegraphics[width=0.45\linewidth]{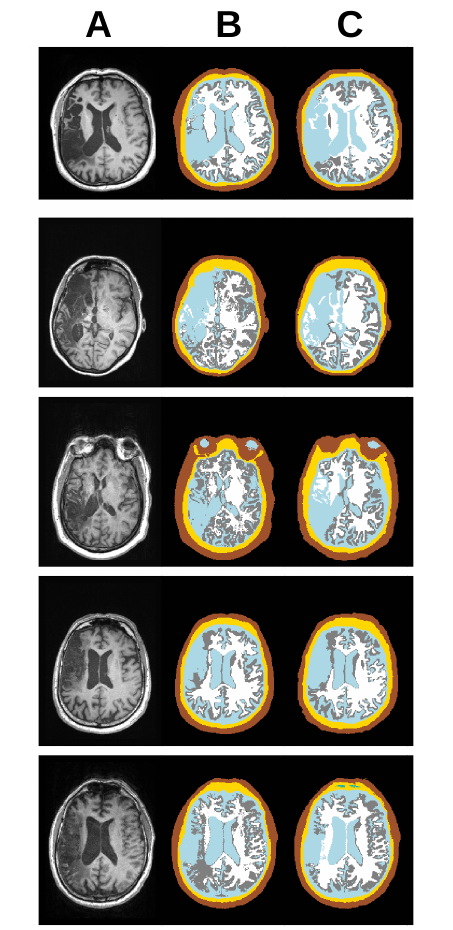}
  \caption{\responseedit{Examples of T1-weighted MRIs (column A) from five different stroke patients and their corresponding segmentation. Manual segmentations done from scratch by two different trainees are displayed in columns B and C. Segmentation colors are arbitrarily chosen to represent the following tissues: CSF = light-blue, white-matter = white, gray-matter = gray, bone = yellow, skin = brown, air = green, background = black.}}
  \label{fig:Stroke_data_segmentations}
\end{figure}


\begin{figure}[ht] 
  \centering
  \includegraphics[width=\linewidth]{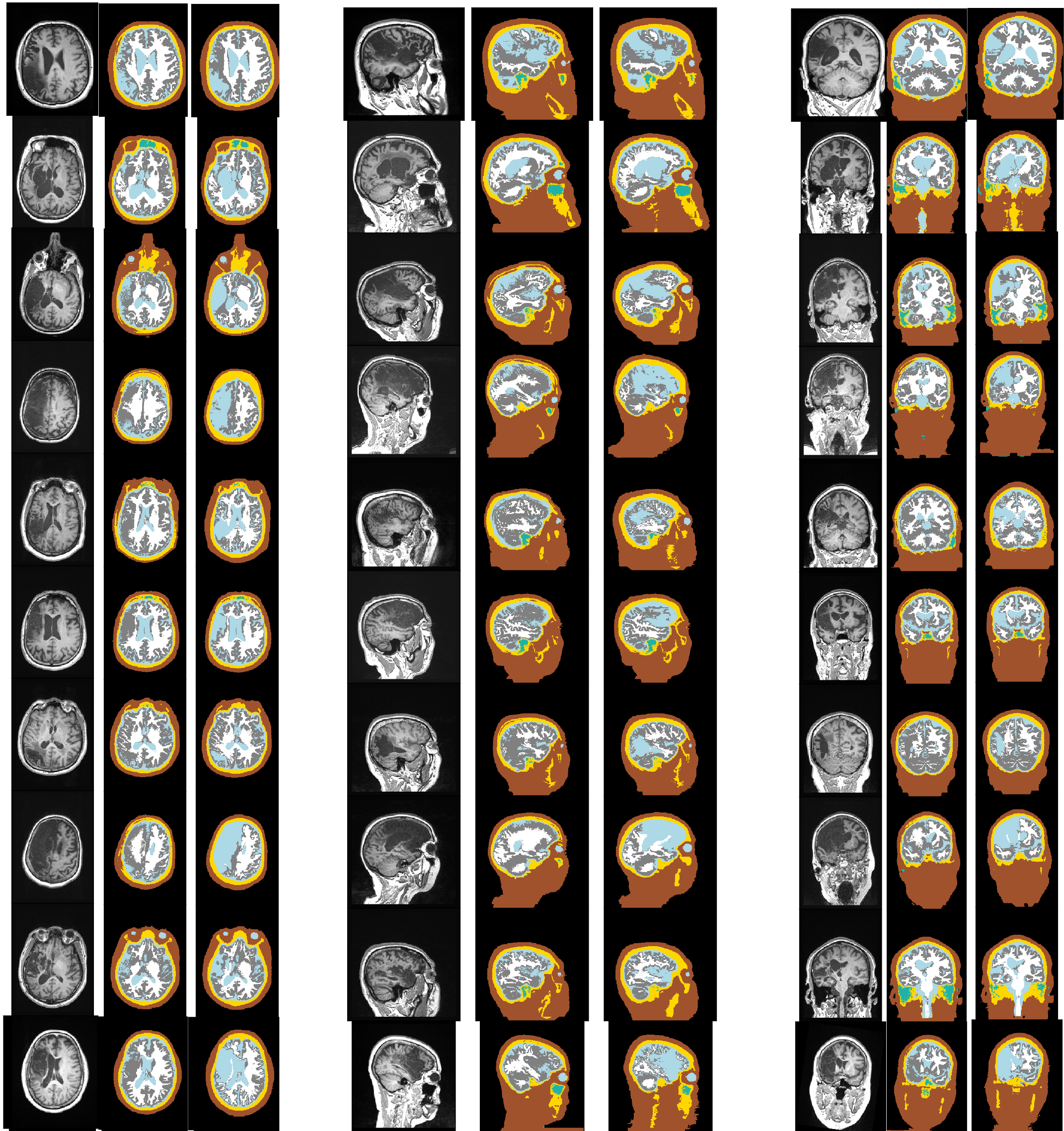}
  \caption{\minoredit{Ten examples of SPM8 segmentations and manual corrections of lesion areas. Each row displays a lesion area in three orthogonal planes (axial, sagittal and coronal in this order) for each subject. For every presented MRI, the corresponding segmentation from SPM8 is shown directly on its right, which often fails to capture the CSF-filled lesion, and right next to it, the segmentation after manual correction, which was used as the target label for the training set. Segmentation colors are arbitrarily chosen to represent the following tissues: CSF = light-blue, white-matter = white, gray-matter = gray, bone = yellow, skin = brown, air = green, background = black.}}
  \label{fig:Stroke_data_segmentations_corrections_part1}
\end{figure}

\begin{figure}[ht] 
  \centering
  \includegraphics[width=\linewidth]{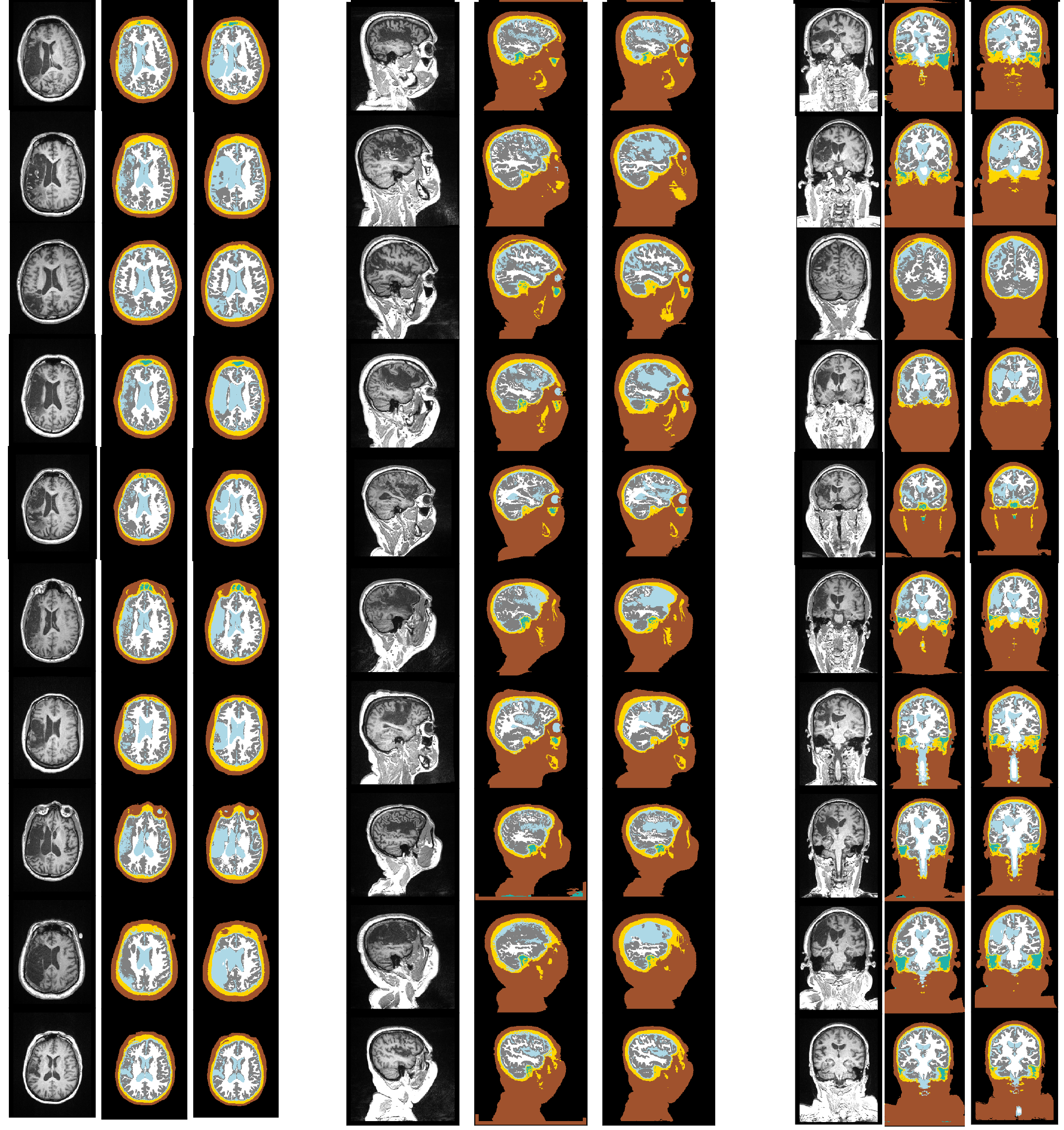}
  \caption{\minoredit{Additional ten examples of SPM8 segmentations and manual corrections of lesion areas. Each row displays a lesion area in three orthogonal planes (axial, sagittal and coronal in this order) for each subject. For every presented MRI, the corresponding segmentation from SPM8 is shown directly on its right, which often fails to capture the CSF-filled lesion, and right next to it, the segmentation after manual correction, which was used as the target label for the training set. Segmentation colors are arbitrarily chosen to represent the following tissues: CSF = light-blue, white-matter = white, gray-matter = gray, bone = yellow, skin = brown, air = green, background = black.}}
  \label{fig:Stroke_data_segmentations_corrections_part2}
\end{figure}

\end{document}